\documentstyle[12pt,epsf]{article}
\begin{document}
\hoffset=-1.2cm
\baselineskip=7.5mm
\hsize=16cm
\vsize=24cm
\begin{titlepage}
\begin{flushright}
ADP-99-8/T353\\
DTP-99/76 \\
July, 1999\\
\end{flushright}
\vskip 4mm
{\centerline{\large{\bf {ANALYTIC FORM OF THE ONE-LOOP VERTEX AND }}}}
%\vskip 2mm
{\centerline{\large{\bf {OF THE TWO-LOOP FERMION PROPAGATOR IN}}}}
{\centerline{\large{\bf {3-DIMENSIONAL MASSLESS QED}}}}
%\vskip 2mm
%{\centerline{(Adelaide University preprint no. ADP-99-8/T353)}}
%{\centerline{(to be submitted to Phys. Rev. D)}}
\vskip 1cm
\baselineskip=6mm
{\centerline{\large{\bf{A. Bashir{$^{1}$}, A. K{\i}z{\i}lers\"{u}{$^{2,3,4}$} and
M.R. Pennington$^{5}$}}}}
\vskip 5mm
{\centerline{{$^{1}$}Instituto de F{\'\i}sica y Matem\'aticas}}
{\centerline{Universidad Michoacana de San Nicol\'as de Hidalgo}}
{\centerline{Apdo. Postal 2-82 , Morelia, Michoac\'an, M\'exico}}
\vskip 0.5cm
{\centerline{{$^2$}Faculty of Science, Physics Department,}}
{\centerline{University of Istanbul, Beyaz{\i}t, Istanbul, Turkey}}
\vskip 0.5cm
{\centerline{{$^3$}Special Research Centre for the Subatomic Structure of Matter,}}
{\centerline{University of Adelaide, Adelaide, Australia 5005}}
\vskip 0.5cm
{\centerline{{$^4$}Department of Physics and Mathematical Physics,}}
{\centerline{University of Adelaide, Adelaide, Australia 5005}}
\vskip 0.5cm
{\centerline{{$^5$}Centre for Particle Theory,}}
{\centerline{University of Durham, Durham DH1 3LE, U.K.}}
\vskip 1cm
{\centerline {ABSTRACT}}
\vskip 3mm
{\leftskip=15mm\rightskip=15mm\noindent We evaluate the analytic expression for the one-loop fermion-boson 
vertex in massless QED3 in an arbitrary covariant gauge. The result is
written in terms of elementary functions of its momenta. The vertex is decomposed
into a longitudinal part, that is fully responsible for ensuring the Ward
and Ward-Takahashi identities are satisfied, and a transverse part. 
Following Ball and Chiu and K{\i}z{\i}lers\"{u} {\it et. al.}, the
transverse part is written in its most general form as a function of 4 
independent vectors. We calculate the coefficients
of each of these vectors. We also check the transversality condition to
two loops and evaluate the fermion propagator to the same order.
We compare our results with a conjecture of the non-perturbative vertex 
by Tjiang and Burden.\par}
\end{titlepage}
\vfil\eject
\vskip 2cm
%\end{titlepage}
\section{Introduction}
\baselineskip=7mm

\indent

     QED in 3-dimensions is a useful laboratory to study Schwinger-Dyson
equations. As compared to its 4-dimensional counterpart, it is relatively
simpler because of the lack of ultraviolet divergences. Moreover, in the
quenched approximation, it exhibits confinement which makes it more 
attractive for non-perturbative studies. The non-perturbative study
of gauge theories through the use of Schwinger-Dyson equations requires
the knowledge of the non-perturbative form of the fundamental 
fermion-boson interaction. The most commonly used approximation is the
bare vertex. However, among other drawbacks, it fails to respect 
a key property of a gauge field theory, namely the gauge invariance 
of physical observables. An obvious reason is that the bare vertex
fails to respect the Ward Takahashi Identity (WTI). Ball and
Chiu~\cite{BC} have proposed an ansatz for what is conventionally
called the longitudinal part of the vertex which alone satisfies WTI. Another
transverse part remains undetermined.    
The QED3 (quenched and unquenched) has been 
studied for dynamical mass generation for fermions in the bare vertex
approximation as well as using an {\em ansatz} for the full vertex 
which is a simple modification of that proposed by Ball and Chiu \cite
{Roberts1, Walsh1}. More recently, another full vertex {\em ansatz} has been 
used to
study fermion and photon propagators simultaneously \cite{Burden1}, including
an explicit transverse piece.

The only truncation of the complete set of Schwinger-Dyson
equations known so far that incorporates the gauge invariance of a gauge 
theory at each level of
approximation is perturbation theory. Therefore, it is natural to
assume that physically meaningful solutions
of the Schwinger-Dyson equations must agree with perturbative results
in the weak coupling regime. It requires, e.g., that every non-perturbative
{\em ansatz} chosen for the full vertex must reduce to its perturbative
counterpart when the interactions are weak. Whereas in QED4 this
realization has been of enormous help to construct physically
acceptable form of the vertex \cite{CP,KRP,BKP}, need exists to
exploit perturbation theory in exploring the non-perturbative form
of the vertex in QED3. Following \cite{KRP}, we perform an analogous 
calculation in 
QED3. We evaluate one-loop vertex in perturbation theory for massless
fermions. Unlike QED4, all the loop integrals involved are perfectly
well-behaved in ultraviolet regime and hence there is no need to
renormalize them. 

The fermion propagator, 
${S_{F}}$, of momentum ${p}$ involves only one function of 
${p^2}$ for massless fermions. We call it ${F(p^2)}$. 
\begin{eqnarray}
{\it i}{\it S}_{F}(p)={\it i}\frac{F(p^2)}{\not\! p}\qquad.
\end{eqnarray}
The Ward-Takahashi identity relates the 3-point 
Greens function to the fermion propagator. The work of Ball and Chiu~\cite{BC}
tells us how to express the non-perturbative structure of the 
 longitudinal part of the vertex in terms of ${F(p^2)}$.
The knowledge of the fermion function ${F(p^2)}$ helps us evaluate 
the longitudinal component of the vertex. The transverse vertex
is obtained by subtracting the longitudinal one from the full
vertex.
In its most general form, the full vertex can be written in terms of 12 
basis tensors.
The Ball-Chiu construction consumes 4 of these to write the longitudinal
vertex and 8 are left to express the transverse part. For massless
fermions, only 4 of the 8 coefficients are non-vanishing.
The vertex should be free of any kinematic singularities.
Ball and Chiu choose the basis in such a way that the coefficient of each
of the basis is independently free of kinematic singularities
in the Feynman gauge.
This basis was later modified to exhibit the same quality
in an arbitrary covariant gauge by K{\i}z{\i}lers\"{u} {\it et. al.} \cite{KRP}.
There is no a priori reason for the coefficients to be free
of kinematic singularities in QED3 as well with the
same choice of basis. However, we find that the same set of basis
vectors serve perfectly well for QED3 as well. We present the final 
expression for all the coefficients in terms of basic functions of the
momenta involved. This result
should serve as a guide in hunting for the non-perturbative form of
the transverse vertex as every such construction should reduce to
it in the weak coupling regime. We also check the transversality
condition to two loops and find that to this order, it is not
realized in perturbation theory. We evaluate $F(p^2)$ to 
${\cal O}(\alpha^2)$ analytically and compare our findings with a recent 
conjecture of
the vertex proposed by Tjiang and Burden.

\section{The Full Vertex}
\subsection{The Non-perturbative Vertex}

   The full vertex, Fig.~1, ${\Gamma^{\mu}(k,p)}$ can be expressed in terms of
12 spin amplitudes formed from the vectors ${\gamma^{\mu},k^{\mu},p^{\mu}}$
and the scalars 1,${\not\!k,\not\!p}$ and ${\not\!k\not\!p}$. Thus we can
write
\begin{eqnarray}
\Gamma^{\mu}=\sum^{12}_{i=1} P^{i}V^{\mu}_{i}   \qquad,
\end{eqnarray}
where we choose the ${V^{\mu}_{\it i}}$ as follows
\begin{eqnarray}
V_{1}^{\mu}&=&{k^{\mu}}{\not \! k}\;\,\,,\,\,\,\,\,\,
V_{2}^{\mu}={p^{\mu}}{\not \! p}\;\;\,,\,\,\,\,\,\,
V_{3}^{\mu}={k^{\mu}}{\not \! p}\,,\,\,\,\,\,\,
V_{4}^{\mu}={p^{\mu}}{\not \! k}\nonumber\\
V_{5}^{\mu}&=&{\gamma^{\mu}}{\not \! k}{\not \! p},\,\,\,\,\,\,
V_{6}^{\mu}={\gamma^{\mu}}\;\;\;\;\,,\,\,\,\,\,\,
V_{7}^{\mu}={k^{\mu}}\,\,\,\;,\,\,\,\,\,\,
V_{8}^{\mu}={p^{\mu}}\nonumber\\
V_{9}^{\mu}&=&{p^{\mu}}{\not \! k}{\not \! p},\,\,\,\,\,\,
V_{10}^{\mu}={k^{\mu}}{\not \! k}{\not \! p},\,\,\,\,\,\,
V_{11}^{\mu}={\gamma^{\mu}}{\not \! k}\,,\,\,\,\,\,\,
V_{12}^{\mu}={\gamma^{\mu}}{\not \! p}\qquad.
\end{eqnarray}
The full vertex satisfies the Ward-Takahashi identity
\begin{eqnarray}
q_{\mu}\Gamma^{\mu}(k,p)={\it S}^{-1}_{F}(k)-{\it S}^{-1}_{F}(p),
\end{eqnarray}
where ${q=k-p}$, and the Ward identity
\begin{eqnarray}
\Gamma^{\mu}(p,p)={\frac{\partial}{\partial p^{\mu}}}{\it S}^{-1}_{F}(p)
\end{eqnarray}
as the non-singular ${k \rightarrow p}$ limit of Eq.~(4). We follow Ball 
and Chiu and define the longitudinal component of the vertex in terms of
the fermion propagator as
\begin{eqnarray}
\Gamma^{\mu}_{L}&=&\frac{\gamma^{\mu}}{2}
\left(\frac{1}{F(k^2)}+\frac{1}{F(p^2)}\right) \; + \; 
\frac{1}{2} \, \frac{({\not \! k}+{\not \! p})(k+p)^{\mu}}
{(k^2-p^2)}\left(\frac{1}{F(k^2)}-\frac{1}{F(p^2)}\right) \;.
\end{eqnarray}
${\Gamma^{\mu}_{L}}$ alone then satisfies the Ward-Takahashi identity,
Eq.~(4), and being free of kinematic singularities the Ward identity,
Eq.~(5), too. The full vertex can then be written as
\begin{eqnarray}
\Gamma^{\mu}(k,p)=\Gamma^{\mu}_{L}(k,p)+\Gamma^{\mu}_{T}(k,p) \qquad,
\end{eqnarray}
where the transverse part satisfies
\begin{eqnarray}
q_{\mu}\Gamma^{\mu}_{T}(k,p)=0\;\;\;\;\;\mbox{and} \;\;\;\;
\Gamma^{\mu}_{T}(p,p)=0\qquad.
\end{eqnarray}
The Ward-Takahashi identity fixes 4 coefficients of the 12 spin amplitudes
in terms of the fermion functions ---  the 3 combinations explicitly
given in Eq.~(6), while the coefficient of ${\sigma_{\mu\nu}k^{\mu}p^{\nu}}$
must be zero~\cite{BC}. The transverse component ${\Gamma^{\mu}_{T}(k,p)}$
thus involves 8 vectors, out of which the following 4 are sufficient to 
describe the
transverse vertex for the case of massless fermions~:
\begin{eqnarray}
\Gamma^{\mu}_{T}(k,p)=\sum_{i=2,3,6,8} \tau_{i}(k^2,p^2,q^2)T^{\mu}_{i}(k,p) 
\qquad,
\end{eqnarray}
where
\begin{eqnarray}
&T^{\mu}_{2}&=\left[p^{\mu}(k\cdot q)-k^{\mu}(p\cdot q)\right]({\not\! k}
+{\not\! p})\nonumber\\
&T^{\mu}_{3}&=q^2\gamma^{\mu}-q^{\mu}{\not \! q}\nonumber\\
&T^{\mu}_{6}&=\gamma^{\mu}(p^2-k^2)+(p+k)^{\mu}{\not \! q}\nonumber\\
&T^{\mu}_{8}&=-\gamma^{\mu}k^{\nu}p^{\lambda}{\sigma_{\nu\lambda}}
+k^{\mu}{\not \! p}-p^{\mu}{\not \! k}\nonumber\\
\mbox{with}\;\;\;\;\;\;\;\;\;
&\sigma_{\mu\nu}&=\frac{1}{2}[\gamma_{\mu},\gamma_{\nu}]\qquad.
\end{eqnarray}
The coefficients ${\tau_{\it i}}$ are Lorentz scalar functions of ${k}$ and
${p}$, i.e., functions of ${k^2,p^2,q^2}$.
\subsection{The one loop calculation}
   The vertex of Fig.~1 can be expressed as
\begin{eqnarray}
\Gamma^{\mu}(k,p)=\,\gamma^{\mu}+\,\Lambda^{\mu}(k,p).
\end{eqnarray}
    Using the Feynman rules, ${\Lambda^{\mu}}$ to
${O(\alpha)}$ is simply given by:
\begin{eqnarray}
-{\it i}e\Lambda^{\mu}\,=\,\int_{M}\frac{d^3w}{(2\,\pi)^3}
(-{\it i}e\gamma^{\alpha}){\it i}{\it S}^{0}_{F}(p-w)(-{\it i}e\gamma^{\mu})
{\it i}{\it S}^{0}_{F}(k-w)(-{\it i}e\gamma^{\beta}){\it i}
\Delta^0_{\alpha\beta}(w) \;,
\end{eqnarray}
where $M$ denotes the loop integral is to be performed in Minkowski space.
The bare quantities are
\begin{eqnarray*}
  -{\it i}e\Gamma^0_{\mu}&=&-{\it i}e\gamma_{\mu}\\
  {\it i}{\it S}^0_{F}(p)&=&{\it i}
  {\not \! p}/{p^2}\\
  {\it i}{\it \Delta}^0_{\mu\nu}(p)&=&-{\it i}\left[p^2
 g_{\mu\nu}+(\xi-1)p_{\mu}p_{\nu}\right]/p^4  \quad,
\end{eqnarray*}
where $e$ is the usual QED coupling and the parameter ${\xi}$ specifies the
covariant gauge.
Substituting these values in Eq.~(12),
we have with ${\alpha \equiv e^2/4\pi}$:
\begin{eqnarray}
\Lambda^{\mu}\,=\,-\frac{{\it i}\,{\alpha}}{2\,{\pi}^2}\int_{M}d^3w\,
\Bigg\{
\frac{A^{\mu}}{w^2\,(p-w)^2\,(k-w)^2}
+(\xi-1)\frac{B^{\mu}}{w^4\,(p-w)^2\,
(k-w)^2}
\Bigg\}
\end{eqnarray}
on separating the ${g_{\alpha\beta}}$ and ${w_{\alpha}w_{\beta}}$
parts of the photon propagator. In the above equation,
\begin{eqnarray}
A^{\mu}\,&=&\,\gamma^{\alpha}\,\left(\,\not \! p-\,\not \! w\right)\,
\gamma^{\mu}\,\left(\,\not \! k-\,\not \! w\right)\,\gamma_{\alpha} 
\qquad, \\
B^{\mu}\,&=&\,\not \! w\,\left(\,\not \! p-\,\not \! w\right)\,
\gamma^{\mu}\,\left(\,\not \! k-\,\not \! w\right)\,\not \! w
\qquad.
\end{eqnarray}
   What makes the present calculation different from the one in 
4-dimensions is that due to the reduction of powers of $w$ in the 
numerator, none of the integrals is ultraviolet divergent.
To proceed we introduce the following seven basic integrals over the
loop momentum ${d^{3}w}$~:
${J^{(0)}, J^{(1)}_{\mu}, J^{(2)}_{\mu\nu}, I_0, I^{(1)}_{\mu},
I^{(2)}_{\mu\nu}}$ and ${K^{(0)}}$.
\begin{eqnarray}
{\it J}^{(0)}&=&\int_{M}\,d^3w\,\frac{1}
{w^2\,(p-w)^2\,(k-w)^2}\\
{\it J}^{(1)}_{\mu}&=&\int_{M}\,d^3w\,\frac{w_{\mu}}
{w^2\,(p-w)^2\,(k-w)^2}\\
{\it J}^{(2)}_{\mu\nu}&=&\int_{M}\,d^3w\,\frac{w_{\mu}w_{\nu}}
{w^2\,(p-w)^2\,(k-w)^2}\\
{\it I}^{(0)}&=&\int_{M}\,d^3w\,\frac{1}
{w^4\,(p-w)^2\,(k-w)^2}\\
{\it I}^{(1)}_{\mu}&=&\int_{M}\,d^3w\,\frac{w_{\mu}}
{w^4\,(p-w)^2\,(k-w)^2}\\
{\it I}^{(2)}_{\mu\nu}&=&\int_{M}\,d^3w\,\frac{w_{\mu}w_{\nu}}
{w^4\,(p-w)^2\,(k-w)^2}\\
{\it K}^{(0)}&=&\int_{M}\,d^3w\,\frac{1}
{(p-w)^2\,(k-w)^2}  \quad.
\end{eqnarray}
${\Lambda^{\mu}}$ of Eq.~(13) can then be re-expressed in terms of five
of these as:
\begin{eqnarray}
\Lambda^{\mu}&=&-\frac{{\it i}\,{\alpha}}{2\,{\pi}^2}
\Bigg\{
\gamma^{\alpha}\,
{\not \! p}\,{\gamma^{\mu}}\,{\not \! k} \, \gamma_{\alpha}
{\bf {\it J}}^{(0)} 
-{\gamma^{\alpha}}\left(
{\not \! p}\,{\gamma^{\mu}}{\gamma^{\nu}}
+{\gamma^{\nu}}{\gamma^{\mu}}{\not \! k} \right)
\gamma_{\alpha}{\bf {{\it J}_{\nu}^{(1)}}}
%\nonumber\\
%
%
%&& \hspace{10mm}
+{\gamma^{\alpha}}{\gamma^{\nu}}{\gamma^{\mu}}{\gamma^{\lambda}}
\gamma_{\alpha}{\bf {{\it J}_{\nu\lambda}^{(2)}}}\nonumber\\
&& \hspace{10mm}
+(\xi-1)\Bigg[
\left(-{\gamma^{\nu}}{\not \! p}\,{\gamma^{\mu}}-\,
{\gamma^{\mu}}{\not \! k}\,{\gamma^{\nu}} \right)
{\bf {\it J}_{\nu}^{(1)}}
+{\gamma^{\mu}}{\bf \it K}^{(0)}
%\nonumber\\
%
%
% && \hspace{25mm}
+{\gamma^{\nu}}{\not \! p}\,{\gamma^{\mu}}{\not \! k}\,{\gamma^{\lambda}}
 {\bf {{\it I}_{\nu\lambda}^{(2)}}}
\Bigg]
\Bigg\}. 
\end{eqnarray}
As the next step, we compute the basic integrals of Eqs.~(16-22),
\cite{tHooft,DD,Davy} each of
which is a function of $k$ and $p$.  

\vspace{3mm}
\noindent
{\underline{\bf{${{\it J}^{(1)}_{\mu}}$ and ${{\it J}^{(2)}_{\mu\nu}}$
calculated:}}}  

\vspace{5mm}
\noindent
Following~\cite{BC,KRP}, we expand the Lorentz vector ${J^{(1)}_{\mu}}$
in its most general form in terms 
of the 4-momenta ${k_{\mu}}$ and ${p_{\mu}}$:
\begin{eqnarray}
{\it J}^{(1)}_{\mu}=\frac{{\it i}\pi^3}{2}\left[
k_{\mu}J_{A}(k,p)+p_{\mu}J_{B}(k,p)\right]
\end{eqnarray}
where ${J_{A}, J_{B}}$ must be scalar functions of ${k}$ and ${p}$.
The factor of
${{\it i}\pi^3/2}$ is taken out purely for later convenience. One can see
from the integral form of ${J^{(1)}_{\mu}}$ that ${J_{B}(k,p)=J_{A}(p,k)}$.
Inverting the above equation,
\begin{eqnarray}
J_{A}(k,p)=\frac{1}{{\it i}\pi^3\Delta^2}\left[2\,k\cdot p\, p^{\mu}J^{(1)}_
{\mu}-2\,p^2k^{\mu}J^{(1)}_{\mu}\right]
\end{eqnarray}
with a similar expression for ${J_{B}}$,
where $\Delta^2=(k\cdot p)^2-k^2p^2$. Moreover,
using the identity
\begin{eqnarray}
2p\cdot w\,=\,p^2+w^2-(p-w)^2
\end{eqnarray}
for any 4-momentum $p$, we can write
\vspace{5mm}
\begin{eqnarray}
 k^{\mu}J^{(1)}_{\mu} &=& \; 
\frac{k^2}{2} \, {\it J}^{(0)} +  \frac{1}{2} \, {\it K}^{(0)}
- \frac{1}{2} \int\frac{d^3w}{w^2 (p-w)^2}
%
%
%\qquad.
\end{eqnarray}
and a similar expression for $p^{\mu}J_{\mu}^{(1)}$. Evaluating the scalar 
integrals involved using dimensional regularization and substituting the result
in Eq.~(25), we arrive at: 
\begin{eqnarray}
J_{A}(k,p)&=&\frac{1}{\Delta^2}   \Bigg\{ 
\frac{p^2 k \cdot q}{\sqrt{-k^2 p^2 q^2}} \; + \; \frac{p \cdot q}{\sqrt{-q^2}} 
\;
-  \; \frac{k \cdot p}{\sqrt{-k^2}}  \; + \; \frac{p^2}{\sqrt{-p^2}} \, 
   \Bigg\}  \\  \nonumber  \\
J_{B}(k,p)&=&J_{A}(p,k)  
\end{eqnarray}
where we have made use of Eqs.~(1,5) of the Appendix.
In an analogous
fashion, the tensor integral ${J^{(2)}_{\mu\nu}}$ of Eq.~(18) can be expressed
in terms of scalar integrals ${K_0}$,${J_{C},J_{D}}$ and ${J_{E}}$ by
\begin{eqnarray}
{\it J}^{(2)}_{\mu\nu}&=&\frac{{\it i}\pi^3}{2}\Bigg\{
\frac{g_{\mu\nu}}{3}K_0+\left(k_{\mu}k_{\nu}-g_{\mu\nu}\frac{k^2}{3}\right)
J_{C}\nonumber\\
&+&\left(p_{\mu}k_{\nu}+k_{\mu}p_{\nu}-g_{\mu\nu}\frac{2(k\cdot p)}{3}\right)
J_{D}+\left(p_{\mu}p_{\nu}-g_{\mu\nu}\frac{p^2}{3}\right)J_{E}
\Bigg\}\qquad,
\end{eqnarray}
where
\begin{eqnarray}
J_{C}(k,p)&=&\frac{1}{2\Delta^2}   \Bigg\{ p^2 (k \cdot p - 2 k^2) \; J_A
 - p^4 \; J_B \; + \; \frac{k \cdot p + p^2}{\sqrt{-q^2}}   \;
-  \; \frac{k \cdot p }{\sqrt{-k^2}}    \Bigg\} \;,  
 \\  \nonumber \\
J_{D}(k,p)&=&\frac{1}{4\Delta^2}   \Bigg\{ k^2 (3 k \cdot p - p^2) \; J_A
 + p^2  (3 k \cdot p - k^2)  J_B \; - \; \frac{(k + p)^2}{\sqrt{-q^2}}  
   \;  \nonumber \\
&& \hspace{15mm} +  \; \frac{k^2}{\sqrt{-k^2}}  \;  +  \; 
\frac{p^2 }{\sqrt{-p^2}}  
   \Bigg\} \;,  
  \nonumber \\
J_{E}(k,p)&=&J_{C}(p,k) \;,
\end{eqnarray}
which involve the previously found $J_{A}$ and $J_{B}$ of 
Eqs.~(28,29).

\vspace{3mm}
\noindent
{\underline{\bf{${{\it I}^{(1)}_{\mu}}$ and ${{\it I}^{(2)}_{\mu\nu}}$
calculated:}}}

\vspace{5mm}
\noindent
In a way analogous to the computation of ${J^{(1)}_{\mu}}$ and
${J^{(2)}_{\mu\nu}}$ the ultraviolet finite integrals ${I^{(1)}_{\mu}}$ and
${I^{(2)}_{\mu\nu}}$~\cite{Davy} of Eqs.~(20,21) can be re-expressed in terms
of
scalar
integrals, ${I_{A}, I_{B}, I_{C}, I_{D}, I_{E}}$, that in turn involve the
same functions we have already computed. Thus
\begin{eqnarray}
{\it I}^{(1)}_{\mu}&=&\frac{{\it i}\pi^3}{2}\left[
 k_{\mu}I_{A}(k,p)+p_{\mu}I_{B}(k,p)\right]  \qquad,
\end{eqnarray}
where
\begin{eqnarray}
I_{A}(k,p)=\frac{-1}{k^2} \, \frac{1}{\sqrt{-k^2 p^2 q^2}}  \hspace{10mm}
{\rm and}
 \hspace{10mm} I_{B}(k,p)=I_{A}(p,k)  \quad.
\nonumber\\
\end{eqnarray}
${\it I}^{(2)}_{\mu\nu}$ can be expressed as
\begin{eqnarray}
{\it I}^{(2)}_{\mu\nu}&=&\frac{{\it i}\pi^3}{2}\Bigg\{
\frac{g_{\mu\nu}}{3}{\it J}_{0}+\left(k_{\mu}k_{\nu}
-g_{\mu\nu}\frac{k^2}{3}\right)I_{C}
\nonumber\\
&& \hspace{10mm}
+\left(p_{\mu}k_{\nu}+k_{\mu}p_{\nu}-g_{\mu\nu}\frac{2(k\cdot p)}{3}\right)I_{D}
+\left(p_{\mu}p_{\nu}-g_{\mu\nu}\frac{p^2}{3}\right)I_{E}
\Bigg\} \;,
\end{eqnarray}
where
\begin{eqnarray}
I_{C}(k,p)&=&\frac{1}{2\Delta^2}   \Bigg\{ p^2 (k \cdot p - 2 k^2) \; I_A
 - p^4 \; I_B \; + (k \cdot p - 2 p^2) \; J_A  - p^2 \; J_B \nonumber \\
 \nonumber \\
 && \hspace{15mm} -\; \frac{1}{\sqrt{-k^2}} \, \frac{k \cdot p}{k^2} \; - \;
   \frac{4p^2}{\sqrt{-k^2 p^2 q^2}}   \Bigg\} \quad,  
\\  \nonumber \\
I_{D}(k,p)&=&\frac{1}{4\Delta^2}   \Bigg\{ k^2 (3 k \cdot p - p^2) \; I_A
 + p^2 (3 k \cdot p - k^2) \; I_B \; + (3 k \cdot p - k^2) \; J_A 
  \nonumber \\ \nonumber \\
 && \hspace{15mm} +\;  
  (3 k \cdot p - p^2) \; J_B 
 \; +\; \frac{1}{\sqrt{-k^2}} \; +  \; \frac{1}{\sqrt{-p^2}} 
 \; + \;  \frac{8k \cdot p}{\sqrt{-k^2p^2q^2}}   \Bigg\} \quad,  
\\  \nonumber \\
I_{E}(k,p)&=&I_{C}(p,k) \quad.
\end{eqnarray}
\noindent
{\underline{\bf{${\Lambda^{\mu}}$ collected:}}}

\vspace{5mm}
\noindent
$\Lambda^{\mu}$ can now be written completely in terms of the basic functions 
${\it J}_0$, ${\it J}_A$, ${\it J}_B$, ${\it J}_C$,
 ${\it J}_D$, ${\it J}_E$, ${\it I}_0$, ${\it I}_A$, ${\it I}_B$, 
 ${\it I}_C$, ${\it I}_D$,
 ${\it I}_E$ and ${K_0}$, all of which depend on the
momenta $k$ and $p$~:
\begin{eqnarray}
\Lambda^{\mu}(k,p)&=&\sum^{6}_{i=1} {\bar P}_1^i\,V^{\mu}_i  \qquad,
\end{eqnarray}
where 
\begin{eqnarray}
{\bar P}_1^i &=& \frac{\alpha}{4} \;  P_1^i 
\end{eqnarray}
and the explicit expressions for $P_1^i$ are:
\begin{eqnarray}
P_1^1&=&2J_{A}-2J_{C}+(\xi-1) \;2 p^2I_{D} \nonumber\\
P_1^2&=&2J_{B}-2J_{E}+(\xi-1) \; 2 k^2I_{D} \nonumber\\
P_1^3&=&-4{\it J}_{0}+4J_{A}+4J_{B}-2J_{D}\nonumber\\
&-& \frac{1}{3} (\xi-1) \left(4 {\it J}_0 - 6 J_A + 2 k^2I_{C}
+\,4 k\cdot p\,I_{D} -\, 4{p^2} I_{E} \right)\nonumber\\
P_1^4&=&2{\it J}_{0}-2J_{A}-2J_{B}-2J_{D}\nonumber\\
&+& \frac{1}{3} (\xi-1) \left(2 {\it J}_0 - 6 J_A + 4 k^2I_{C}
-\,4 k\cdot p\,I_{D} -\, 2{p^2} I_{E} \right)\nonumber\\
P_1^5&=&3(J_{0}-J_{A}-J_{B}) + (\xi-1) \left( J_{0}-J_{A}-J_{B} 
\right)\nonumber\\
P_1^6&=&  \frac{1}{3} \big( -6 k\cdot p\, {\it J}_{0} + 3
( 2 k\cdot p\, - k^2) J_{A} + 3 ( 2 k\cdot p\, - p^2) J_{B}  \nonumber \\
&& \hspace{5mm}+ K_0 + 2 k^2 J_C + 4 k\cdot p\, J_D + 2 p^2 J_E \big) \nonumber\\
&+& \frac{1}{3} (\xi-1) \big(    -2 k\cdot p\, {\it J}_{0}
- 3 k^2 J_{A} - 3 p^2 J_{B}  \nonumber   \\
&& \hspace{5mm} + 3 K_0 + 2 k^2 k\cdot p\,  I_C + 4 (k\cdot p)^2 I_D + 2 p^2 k\cdot p 
I_E  \big) \;.  
\end{eqnarray}
This is the complete one loop correction to the
QED3 vertex in any covariant gauge for massless fermions.
%
%
%
%
%\vfil\eject
%
\section{Analytic Structure of the Vertex}
\subsection{The Longitudinal vertex}
\noindent
{\underline{\bf{$F(p^2)$ in perturbation theory:}}}

\vspace{5mm}
\noindent
As explained in Sect.~2.1, owing to the Ward-Takahashi identity, the 
longitudinal component of the vertex is determined by the fermion 
function, ${F(p^2)}$. In perturbation theory to order ${\cal O}(\alpha)$,
one has to evaluate the graph in Fig.~2. The corresponding mathematical
equation is:
\begin{eqnarray}
  i S_F^{-1}(p) &=& i {S_F^0}^{-1}(p) + e^2 \; \int \frac{d^3k}{(2 \pi)^3}
  \; \gamma^{\mu} \,  S_F^0(k) \, \gamma^{\nu} \, \Delta_{\mu \nu}^0(q)
  \qquad,
\end{eqnarray}
The photon propagator can be split into the transverse and the longitudinal
parts as:
\begin{eqnarray}
     \Delta_{\mu \nu}^0(q) &=&  {\Delta_{\mu \nu}^0}^T(q) - \xi \; 
     \frac{q_{\mu}q_{\nu}}{q^4} \qquad,
\end{eqnarray}
where
\begin{eqnarray}
      {\Delta_{\mu \nu}^0}^T(q) &=& -\frac{1}{q^2} \; \left[
 g_{\mu\nu}- q_{\mu}q_{\nu}/q^2\right] \qquad.
\end{eqnarray}
Burden and Roberts (see Eq. (25) of \cite{trans}) have noted that the solution of Eq. (42)
is gauge covariant (in the sense of the Landau-Khalatnikov transformations
\cite{LK}) if the condition
\begin{eqnarray}
\int \frac{d^3k}{(2 \pi)^3}
  \; \gamma^{\mu} \,  S_F(k) \, \Gamma^{\nu}(k,p) \, 
  {\Delta_{\mu \nu}^0}^T(q)&=&0
\end{eqnarray}
is simply satisfied. This condition Burden and Tjiang \cite{Burden1}
have called the {\it transversality condition}. It is easy to check
that at one loop order this condition is indeed fulfilled and so we are left
with
\begin{eqnarray}
 i S_F^{-1}(p) &=& i {S_F^0}^{-1}(p) + e^2 \; \int \frac{d^3k}{(2 \pi)^3}
  \; \gamma^{\mu} \,  S_F^0(k) \, \gamma^{\nu} \, \left( - \xi \; 
     \frac{q_{\mu}q_{\nu}}{q^4} \right) \qquad.
\end{eqnarray}
Substituting the values of $S_F(p)$ and ${S_F^0}(p)$, then taking the trace after
having multiplied with ${\not \! p}$ and simplifying, we get
\begin{eqnarray}
  \frac{1}{F(p^2)} &=& 1 - i \xi \, \frac{e^2}{p^2} \, 
  \int \frac{d^3k}{(2 \pi)^3} \; \frac{k \cdot p}{k^2 q^2}  \qquad,
\end{eqnarray}
which can also be written as
\begin{eqnarray}
 \frac{1}{F(p^2)} &=& 1 - i \xi \, \frac{e^2}{2p^2} \, 
  \int \frac{d^3k}{(2 \pi)^3} \; \frac{k^2+p^2-q^2}{k^2 q^2} \qquad.
\end{eqnarray}
Using dimensional regularization, one can see that the last term
(and also the first term after appropriate change of variables)
is zero as there are no external momenta present in the integrand.
Eq.~(1) in the appendix simplifies the result to
\begin{eqnarray}
 F(p^2)&=&1-\frac{\alpha\xi}{4} \, \frac{\pi}{\sqrt{-p^2}} \; + \;
  {\cal O}({\alpha}^2) 
\end{eqnarray}
in the Minkowski space. So the longitudinal vertex to ${\cal O}(\alpha)$ 
is:
\begin{eqnarray}
    \Gamma^{\mu}_{L} &=& \left[ 1 + \frac{\alpha \xi}{4} \, \sigma_1 \right]
    \, \gamma^{\mu} \; + \;  \frac{\alpha \xi}{4} \, \sigma_2 \, \left[ 
{k^{\mu}}{\not \! k} \, +  \, {p^{\mu}}{\not \! p} \, + \,
{k^{\mu}}{\not \! p} \, +  \, {p^{\mu}}{\not \! k}  \right] \qquad,
\end{eqnarray}
where
\begin{eqnarray}
   \sigma_1 &=& \frac{1}{2} \, \left[ \frac{\pi}{\sqrt{-k^2}} \, + \, 
\frac{\pi}{\sqrt{-p^2}}  \right]  \qquad, \nonumber \\
   \sigma_2 &=& \frac{1}{2} \, \frac{1}{(k^2-p^2)} \, 
\left[ \frac{\pi}{\sqrt{-k^2}} \, - \, 
\frac{\pi}{\sqrt{-p^2}} \right]  \qquad.
\end{eqnarray}
{\underline{\bf{Comparison with LKF transformations:}}}

\vspace{5mm}
\noindent
Assuming that $F(p^2)=1$ in the Landau gauge, LKF transformations yield
the following expression for it in an arbitrary gauge: 
\begin{eqnarray}
F(p^2)&=&1-\frac{\alpha\xi}{2 \sqrt{-p^2}} \; 
{\rm tan}^{-1} \left[ \frac{2 \sqrt{-p^2}}{\alpha\xi} \right] \qquad.
\end{eqnarray}
Using the expansion ${\rm tan}^{-1}(1/x)= \pi/2 - x + 
x^3 /3 + \cdots$ for $ \mid x \mid << 1$, we  get
\begin{eqnarray}
   F(p^2)&=& 1 - \frac{\pi \, \alpha \xi}{4 \sqrt{-p^2}} - \frac{\alpha^2 \xi^2}{4p^2}
               + {\cal O}(\alpha^3)
\end{eqnarray}
which is in accordance with the perturbative result to ${\cal O}(\alpha)$.
Therefore, the LKF transformations accompanied by the assumption that
$F(p^2)=1$ in the Landau gauge are in accordance with perturbation theory
at the one loop level. A similar comparison at the too loop level
is discussed in Sect. 4.

\noindent
{\underline{\bf{Remark:}}}

\vspace{5mm}
\noindent
Burden and Tjiang \cite{Burden1} propose the following non-perturbative
expression for $F(p^2)$:
\begin{eqnarray}
F(p^2)&=&1-\frac{\alpha(\xi - \xi_0)}{2 \sqrt{-p^2}} \; 
{\rm tan}^{-1} \left[ \frac{2 \sqrt{-p^2}}{\alpha(\xi-\xi_0)} \right] \qquad.
\end{eqnarray}
From the arguments given above, it is easy to see that this expression
agrees with the one loop perturbative result only if $\xi_0=0$, unlike
what is suggested by them.

\subsection{The Transverse Vertex}
Having calculated the vertex to ${O(\alpha)}$ , Eq.~(11,12,39-41), we can subtract 
from
it the longitudinal vertex of Sect.~3.1, Eq.~(51,52) and obtain (Eq.~(9)) the
transverse vertex to ${O(\alpha)}$. This is given by
%
%
%
%\newpage
\begin{eqnarray}
\Gamma^{\mu}_{T}(k,p)&=&\frac{\alpha}{4} \;
\Bigg[ \hspace{4mm} {k^{\mu}}{\not \! k} \hspace{4mm} \{ \, 2J_{A}-2J_{C}
- \sigma_2 \, + \, 
(\xi-1) \; ( 2 p^2I_{D} - \sigma_2 ) \, \}
\nonumber\\
&& \hspace{7mm} + \, {p^{\mu}}{\not \! p} \hspace{4mm} \{ \, 
2J_{B}-2J_{E}- \sigma_2 \, + \, (\xi-1) \; ( 2 k^2I_{D} - \sigma_2) 
 \, \}
\nonumber\\
&& \hspace{7mm} + \, {k^{\mu}}{\not \! p} \hspace{4mm} \{ \, 
-4{\it J}_{0}+4J_{A}+4J_{B}-2J_{D} - \sigma_2 \nonumber\\
&& \hspace{23mm} - \frac{1}{3} \, (\xi-1) \;  \left(4 {\it J}_0 - 6 J_A 
+ 2 k^2I_{C}
+\,4 k\cdot p\,I_{D} -\, 4{p^2} I_{E} + 3 \sigma_2 \right) 
 \, \}
\nonumber\\
&& \hspace{7mm} + \, {p^{\mu}}{\not \! k} \hspace{4mm} \{ \, 
2{\it J}_{0}-2J_{A}-2J_{B}-2J_{D} - \sigma_2 \nonumber\\
&& \hspace{23mm} + \frac{1}{3} \, (\xi-1) \; \left(2 {\it J}_0 - 6 J_A + 4 k^2I_{C}
-\,4 k\cdot p\,I_{D} -\, 2{p^2} I_{E} - 3 \sigma_2  \right)
 \, \}
 \nonumber \\
&& \hspace{7mm} + \, {\gamma^{\mu}}{\not \! k}{\not \! p} \; \{ \,  
3 \, (J_{0}-J_{A}-J_{B}) + (\xi-1) \; \left( J_{0}-J_{A}-J_{B} 
\right)
\, \}
\nonumber\\
&& \hspace{7mm} + \, {\gamma^{\mu}}  \hspace{7mm} \{ \,
\frac{1}{3} \, \big( -6 k\cdot p\, {\it J}_{0} + 3
( 2 k\cdot p\, - k^2) J_{A} + 3 ( 2 k\cdot p\, - p^2) J_{B}  \nonumber \\
&& \hspace{30mm}+ K_0 + 2 k^2 J_C + 4 k\cdot p\, J_D + 2 p^2 J_E 
- 3 \sigma_1 \big) \nonumber\\
&&\hspace{23mm}   + \frac{1}{3} \, (\xi-1) \; \big(    -2 k\cdot p\, {\it J}_{0}
- 3 k^2 J_{A} - 3 p^2 J_{B}  \nonumber   \\
&& \hspace{28mm} + 3 K_0 + 2 k^2 k\cdot p\,  I_C + 4 (k\cdot p)^2 I_D + 2 p^2 k\cdot p 
I_E - 3 \sigma_1 \big)
\, \}  \Bigg]
\end{eqnarray}
in terms of 6 of the vectors ${V^{\mu}_{{\it i}}}$.
Our task is then to express this result in terms of the 4 basis vectors
defining ${\Gamma^{\mu}_{T}(k,p)}$, Eq.~(10). Thus
from Eq.~(9) we can alternatively write out
\begin{eqnarray}
\Gamma^{\mu}_{T}&=&k^{\mu}{\not \! k}  \hspace{2.5 mm} \left[\tau_{2}(p^2-k\cdot p)-\tau_{3}
+\tau_{6}\right]\nonumber\\
&+&p^{\mu}{\not \! p} \hspace{3 mm} \left[\tau_{2}(k^2-k\cdot p)-\tau_{3}-\tau_{6}\right]
\nonumber\\
&+&k^{\mu}{\not \! p} \hspace{3 mm} \left[\tau_{2}(p^2-k\cdot p)+\tau_{3}-\tau_{6}
+\tau_{8}\right]\nonumber\\
&+&p^{\mu}{\not \! k} \hspace{3 mm} \left[\tau_{2}(k^2-k\cdot p)+\tau_{3}
+\tau_{6}-\tau_{8}\right]\nonumber\\
&+&\gamma^{\mu}{\not \! k}{\not \! p} \hspace{1 mm} \left[-\tau_{8}\right]
\nonumber\\
&+&\gamma^{\mu} \hspace{6 mm}  \left[\tau_{3}q^2+\tau_{6}(p^2-k^2)+\tau_{8}(k\cdot p)\right]
\qquad.
\end{eqnarray}
Comparing Eqs.~(55) and (56), we have 6 equations for the 4 unknown
${\tau_{i}}$. Since ${\Gamma^{\mu}_{T}}$ is transverse to the vector
${q_{\mu}}$, Eq.~(4), only 4 of these equations are independent.
The solution yields expressions for the 4 transverse coefficients
${\tau_{i}}$. Each is a function of ${k^2, p^2, q^2}$ and ${\xi}$. The
results are as follows:
\newpage
\newpage
\begin{eqnarray}
\tau_{2}&=&\frac{\alpha \pi}{8\Delta^4} \hspace{10mm} \Bigg\{ 
 \frac {1}{\sqrt{-k^2p^2q^2}} \;  q^2  \left[ (k \cdot p)^2 + k^2 p^2 \right] 
\nonumber\\
&& \hspace{20mm} +  \frac{1}{\sqrt{-k^2}} \; \frac{1}{(k^2-p^2)}
\left[ \Delta^2 (p^2+\,k\cdot p\,) - 2 k\cdot p\ (k^2-p^2) (k^2-\,k\cdot p\,)
\right]  \nonumber \\
&& \hspace{20mm} -  \frac{1}{\sqrt{-p^2}} \; \frac{1}{(k^2-p^2)}
\left[ \Delta^2 (k^2+\,k\cdot p\,) + 2 k\cdot p\ (k^2-p^2) (p^2-\,k\cdot p\,)
\right]  \nonumber \\
&& \hspace{20mm} +  \frac{1}{\sqrt{-q^2}} \; 2 q^2 k\cdot p  \Bigg\} \nonumber \\
&+&\frac{\alpha \pi (\xi-1) }{8\Delta^4} \; \; \Bigg\{ 
-\frac {1}{\sqrt{-k^2p^2q^2}} \; \left[ (k^2 + p^2) \Delta^2 + 2 k^2 p^2 q^2 \right]  
\nonumber\\
&& \hspace{20mm} +  \frac{1}{\sqrt{-k^2}} \; \frac{1}{(k^2-p^2)}
\left[ \Delta^2 (p^2+\,k\cdot p\,) - 2 k^2 (k^2-p^2) (p^2-\,k\cdot p\,)
\right]  \nonumber \\
&& \hspace{20mm} -  \frac{1}{\sqrt{-p^2}} \; \frac{1}{(k^2-p^2)}
\left[ \Delta^2 (k^2+\,k\cdot p\,) + 2 p^2 (k^2-p^2) (k^2-\,k\cdot p\,)
\right]  \nonumber \\
&& \hspace{20mm} -  \frac{1}{\sqrt{-q^2}} \; 2 \left[ q^2 k\cdot p + \Delta^2
\right] \Bigg\} \qquad, \\  \nonumber \\  \nonumber \\
\tau_{3}&=&\frac{\alpha \pi}{16\Delta^4} \hspace{9mm} \Bigg\{ 
-\frac {1}{\sqrt{-k^2p^2q^2}} \;  \left[ -4 (k \cdot p)^2 \Delta^2 + (k^2-p^2)^2
\left( (k \cdot p)^2 + k^2 p^2 \right)  \right]
\nonumber\\
&& \hspace{20mm} +  \frac{1}{\sqrt{-k^2}} \;
\left[ \Delta^2 (p^2-\,k\cdot p\,) + 2 k\cdot p\ (k^2-p^2) (k^2+\,k\cdot p\,)
\right]  \nonumber \\
&& \hspace{20mm} +  \frac{1}{\sqrt{-p^2}} \; 
\left[ \Delta^2 (k^2-\,k\cdot p\,) - 2 k\cdot p\ (k^2-p^2) (p^2+\,k\cdot p\,)
\right]  \nonumber \\
&& \hspace{20mm} +  \frac{1}{\sqrt{-q^2}} \; 
\left[  2 k\cdot p \left( 2 \Delta^2 - (k^2 - p^2)^2 \right) \right]
 \Bigg\} \nonumber \\
&+&\frac{\alpha \pi (\xi-1) }{16\Delta^4} \; \; \Bigg\{ 
-\frac {1}{\sqrt{-k^2p^2q^2}} \; \left[  \Delta^2 (k^2 + p^2)^2 - 2 (k \cdot p)^2
(k^2 - p^2)^2  \right]  
\nonumber\\
&& \hspace{20mm} +  \frac{1}{\sqrt{-k^2}} \; 
\left[ (\Delta^2 + 2 k^2 p^2) (p^2+\,k\cdot p\,) - 2 k^2 k \cdot p 
(k^2+\,k\cdot p\,)
\right]  \nonumber \\
&& \hspace{20mm} +  \frac{1}{\sqrt{-p^2}} \; 
\left[ (\Delta^2 + 2 k^2 p^2) (k^2+\,k\cdot p\,) - 2 p^2 k \cdot p 
(p^2+\,k\cdot p\,)
\right]  \nonumber \\
&& \hspace{20mm} +  \frac{1}{\sqrt{-q^2}} \; \left[ -2 k \cdot p
\left( 2 \Delta^2 - (k^2 - p^2)^2 \right)
\right] \Bigg\}  \qquad,
\end{eqnarray}
\newpage
\begin{eqnarray}
\tau_{6}&=&\frac{\alpha \pi(\xi-2) }{16\Delta^4} \; \; \Bigg\{ 
-\frac {1}{\sqrt{-k^2p^2q^2}} \; \left[ (p^2 - k^2) q^2  \left( (k \cdot p)^2
+ k^2 p^2 \right)   \right]  
\nonumber\\
&& \hspace{20mm} -  \frac{1}{\sqrt{-k^2}} \; 
\left[  \Delta^2 (p^2-\,k\cdot p\,) + 2 k^2 \left( p^2 (p^2 - k \cdot p)
+ k\cdot p  (k^2 - k \cdot p)  \right)
\right]  \nonumber \\
&& \hspace{20mm} -  \frac{1}{\sqrt{-p^2}} \; 
\left[  \Delta^2 (\,k\cdot p\,-k^2) - 2 p^2 \left( k^2 (k^2 - k \cdot p)
+ k\cdot p  (p^2 - k \cdot p)  \right)
\right]   \nonumber \\
&& \hspace{20mm} +  \frac{1}{\sqrt{-q^2}} \; \left[ 2 k \cdot p \;  q^2
(k^2 - p^2)
\right] \Bigg\} \qquad, \\  \nonumber \\  \nonumber \\
\tau_{8}&=&\frac{\alpha \pi (\xi+2) }{4\Delta^2} \; \; \Bigg\{ 
\frac{-k \cdot p \, q^2 }{\sqrt{-k^2p^2q^2}} \;  
+    \frac{k^2 - k \cdot p}{\sqrt{-k^2}} 
+  \frac{p^2 - k \cdot p}{\sqrt{-p^2}} 
- \frac{q^2}{\sqrt{-q^2}}
 \Bigg\} \qquad .
\end{eqnarray}
These ${\tau_i}$ are given in an arbitrary covariant gauge specified by
${\xi}$, written in the Minkowski space. 
  Any non-perturbative vertex {\em ansatz} should reproduce Eqs.~(57-60) in the 
weak coupling regime. Therefore, Eqs.~(57-60) should serve as 
a guide to constructing non-perturbative vertex in QED3.
\begin{itemize}

\item 
   The ${\tau_i}$ have the required symmetry under the exchange of
vectors $k$ and $p$. ${\tau_2}$, ${\tau_3}$ and ${\tau_8}$ are symmetric,
whereas ${\tau_6}$ is antisymmetric.

\item
   None of the ${\tau_i}$ has kinematic singularity when $k^2 \to p^2$.
Although ${\tau_2}$ has explicit factors of $(k^2-p^2)$ in the denominator,
the terms containing them obviously cancel out in the limit $k^2 \to p^2$.

\item
   All the $\tau_i$ only depend on basic functions of $k$ and $p$. This
is unlike the case of QED4 where the $\tau_i$ involve spence functions.

\end{itemize}
It is instructive to take the asymptotic limit $ \mid k^2 \mid >> 
\mid p^2 \mid $ of the transverse 
vertex, as another check of the correctness of Eqs.~(57-60):
\begin{eqnarray}
\nonumber \\
    \tau_2  &\stackrel{ \mid k^2 \mid >> \mid p^2 \mid} {=}& - \frac{\alpha}{16k^4} \; 
           \frac{\pi}{\sqrt{-p^2}} \; (2-3 \xi) \; + \; {\cal O}(1/k^5)  \\
	   \nonumber \\
    \tau_3  &\stackrel{\mid k^2 \mid>> \mid p^2 \mid}{=}& - \frac{\alpha}{32k^2} \; 
           \frac{\pi}{\sqrt{-p^2}} \; (2+3 \xi) \; + \; {\cal O}(1/k^3)  \\
	   \nonumber  \\
    \tau_6  &\stackrel{\mid k^2 >> \mid p^2 \mid}{=}& - \frac{\alpha}{32k^2} \; 
           \frac{\pi}{\sqrt{-p^2}} \; (2- \xi) \; + \; {\cal O}(1/k^3)  \\
	   \nonumber \\ 
    \tau_8  &\stackrel{\mid k^2 >> \mid p^2 \mid }{=}& - \frac{\alpha}{4k^2} \; 
           \frac{\pi}{\sqrt{-p^2}} \; (2+ \xi) \; + \; {\cal O}(1/k^3)  
	   \;.
\end{eqnarray}
Note that 

\begin{itemize}

\item the factors of $\Delta^2$ in each of the $\tau_i$ cancel out
and there is no dependence on the angle between $k$ and $p$, as
expected.

\item taking into account the asymptotic limit  $\mid k^2 \mid>> \mid p^2
\mid $ of the 
corresponding
basis vectors, one can easily see that $\tau_3$ and $\tau_6$
provide the dominant contribution to $\Gamma_T$ in this
limit as in QED4.

\end{itemize}
Therefore, the complete transverse vertex in the limit $\mid k^2 \mid >> \mid
p^2 \mid $
can be written as
\begin{eqnarray}
\nonumber \\
   \Gamma^{\mu}_T(k,p) \stackrel{\mid k^2 \mid >> \mid p^2\mid }{=} \frac{\alpha \xi}{8}
   \; \frac{\pi}{\sqrt{-p^2}} \; \left[ - \gamma^{\mu} + 
   \frac{k^{\mu}{\not \! k}}{k^2}  \right] \qquad. \\ \nonumber
\end{eqnarray}
This result is strikingly similar to that found in QED4, apart from
a factor of ${\rm ln}(k^2/p^2)$ replaced with $\pi/\sqrt{-p^2}$. Note
that in this limit, the exact QED3 vertex matches onto the proposed 
Curtis-Pennington vertex \cite{CP}.

\section{$F(p^2)$ to Two Loops and Transversality condition}
\subsection{$F(p^2)$ to Two Loops}

   We have seen that the transversality condition, i.e, Eq.~(45), holds
true to one loop level. Assuming it to be true non-perturbatively for
$\xi=\xi_0$, which we have shown to be equal to zero, Burden
{\it et. al.} \cite{Burden1}, have proposed a vertex anstaz, which they then
use to solve the photon propagator equation. A crucial test of the
validity of their vertex ansatz is checking the transversality condition
to two loop order. This is equivalent to calculating $F(p^2)$ to the 
same level. We carry out this exercise in this section. 

   The equation for
$F(p^2)$ can be extracted from Eq.~(42) by multiplying the equation with
${\not \! p}$ and taking the trace. On Wick rotating to the Euclidean
space and simplifying, this equation can be written as:
\newpage
\begin{eqnarray}
\nonumber
\frac{1}{F(p^2)}&=&1 \;-\; \frac{\alpha}{2 \pi^2 p^2} \; \int 
\frac{d^3k}{k^2} \frac{F(k^2)}{q^2}  \\ \nonumber
&&  \hspace{10mm} \Bigg[a(k^2,p^2) \, \frac{2}{q^2}  \, \left\{ ( k \cdot p )^2
 - (k^2+p^2) k \cdot p +k^2 p^2 \right\}    \\ \nonumber
&&    \hspace{13mm} + b(k^2,p^2)   \, \left\{ (k^2+p^2) k \cdot p +2 k^2 p^2
      - \frac{1}{q^2} (k^2-p^2)^2 k \cdot p  \right\} \\ \nonumber
&&  \hspace{13mm} - \frac{\xi}{F(p^2)} \frac{1}{q^2}   \, 
\left\{ p^2 (k^2 - k \cdot p) \right\} \\ \nonumber
&&  \hspace{13mm} + \tau_2(k,p) \; \;\left\{-(k^2+p^2) \Delta^2  \right\}
   \\ \nonumber
&&   \hspace{13mm} +  \tau_3(k,p)\, 2 \left\{- (k \cdot p)^2 
   + (k^2+p^2) k \cdot p - k^2 p^2  \right\}   \\ \nonumber
&&  \hspace{13mm} -  \tau_6(k,p) \, 2 \left\{ (k^2-p^2) k \cdot p  \right\} \\
&& \hspace{13mm} +  \tau_8(k,p) \; \; \,\left\{ \Delta^2 \right\}
\Bigg]   \qquad,
\end{eqnarray}
where
\begin{eqnarray}
a(k^2,p^2)= \frac{1}{2} \, \left( \frac{1}{F(k^2)} + 
\frac{1}{F(p^2)} \right),
\hspace{7mm} 
b(k^2,p^2)= \frac{1}{2} \frac{1}{k^2-p^2} \, 
\left( \frac{1}{F(k^2)} - \frac{1}{F(p^2)} \right) \;.
\end{eqnarray}
In connection with carrying out the integral in the above equation,
it is convenient to write $\tau_i$, Eq.~(57-60), in the Euclidean space,
and adopt the notation
$k=\sqrt{k^2}$, $p=\sqrt{p^2}$,  $q=\sqrt{q^2}$. The only angular dependence
is hence displayed in $q= \sqrt{k^2 + p^2 - 2 kp \rm{cos \theta}}$:
\begin{eqnarray}
  \tau_2 &=& \frac{\alpha \pi}{4} \; \frac{1}{kp(k+p)(k+p+q)^2} \;
 \left[  1 + (\xi-1) \, \frac{2k+2p+q}{q}  \right] \;, \\ \nonumber \\
   \tau_3 &=& \frac{\alpha \pi}{8} \; \frac{1}{kpq(k+p+q)^2} \;
 \left[ 4kp+3kq+3pq+2q^2 + (\xi-1) \, (2k^2+2p^2+kq+pq)  \right] , \nonumber\\
   \\
 \tau_6 &=& \frac{\alpha \pi (2- \xi)}{8} \; \frac{k-p}{kp(k+p+q)^2} \;, \\ 
   \nonumber \\
 \tau_8 &=& \frac{\alpha \pi (2+ \xi)}{2} \; \frac{1}{kp(k+p+q)} \;.\\ \nonumber
\end{eqnarray}
We substitue Eqs.~(68-71) in Eq.~(66). Again 
employing the standard technique to identify $ 2  k \cdot p =  (k^2 + p^2 -q^2)
 $ and making use of $d^3k=2 \pi dk k^2 d {\rm \theta sin \theta}$,
we carry out the angular integration:
\newpage
\begin{eqnarray}
\nonumber
    \frac{1}{F(p^2)} &=& 1 + \frac{\pi \xi}{4 p} \, \alpha - \frac{\alpha^2}{4 p^2} 
    \int_0^{\infty} dk  \, \frac{1}{2kp(k+p)} \\ \nonumber
 && \hspace{-5mm}  \Bigg[ \frac{\xi}{2} \, (k^2-p^2) \, \left\{ 
   -(k^2-p^2)^2 I_4 + I_0 \right\}   \\ \nonumber
&& \hspace{-2mm} + \frac{\xi}{2} \, \left\{ (k^2-p^2)^2 (k^2+p^2) I_4 -
    2 (k^2+p^2)^2 I_2 + (k^2 + p^2) I_0 \right\}   \\ \nonumber
&& \hspace{-2mm} -\xi^2 p^2 (k^2-p^2)  \left\{ (k^2 - p^2) I_4 + I_2  \right\}
    \\ \nonumber
&& \hspace{-2mm} + \big\{ (k+p) \left(2kp(k-p)^2 I_3 + (k-p)^2 (k+p) I_2
            - (k^2+p^2) I_1 - (k+p) I_0 + I_{-1} \right) \\ \nonumber
&& \hspace{1mm} +\xi \big( (k-p)^2 (2k^2+2p^2+3kp) I_2 - (k+p) (k^2+p^2-4kp)
       I_1    \\  
       && \hspace{1mm} - (2k^2+2p^2 + 3kp) I_0 
  + (k+p) I_{-1} \big) \big\}    
      \Bigg]    \quad,
\end{eqnarray}
where
\begin{itemize}

\item 

    the first curly-bracket expression arises from the $a$--term in Eq.~(65), 
the second one from the $b$--term, the third from the $\xi/F(p^2)$-term
and the fourth from the transverse part of the vertex. On substituting
$I_4$, $a$--term vanishes identically as it does at one loop level.
Note that
all the $(k+p+q)$ factors in the $\tau_i$ neatly cancel
out, leaving us with simpler integrals to be evaluated.

\item 

and the $I_n$ are defined as
\begin{eqnarray*}
    I_n &=& \int_0^{\pi} \, d \theta \; \frac{{\rm sin} \theta}{q^n}
\end{eqnarray*}
with the evaluated expressions given in the appendix.

\end{itemize}
Keeping in mind the form of the integrals $I_n$, we divide the integration
region in two parts, $0 \rightarrow p$ and $p \rightarrow \infty$. For the
first region, we make the change of variables $k=px$ and for the
second region, $k=p/x$. On simplification, we arrive at
\begin{eqnarray}
\nonumber 
  \frac{1}{F(p^2)} &=& 1 + \frac{\pi \xi}{4 p} \alpha + 
  \frac{\alpha^2 \xi^2}{8 p^2} \int_0^1 \frac{dx}{x} \; \left[ 2 - (1-x)^2
   {\cal L} \right]  \\  \nonumber
 &&- \frac{\alpha^2}{24p^2} \int_0^1 \frac{dx}{x^2} \left[ -4x^2 (x+1)
      -6 (3x^2+1) + 3(1-x^2)^2 {\cal L}  \right] \\
 && - \frac{\alpha^2 \xi}{8 p^2} \int_0^1 \frac{dx}{x^2} \left[ 
  - \frac{2}{3} (2x-1)(x^2-3x-3)+(1+x)^2 (x^2-3x+1) {\cal L} \right] \,,
\end{eqnarray}
where
\begin{eqnarray}
       {\cal L} &=& \frac{1}{x} \; {\rm ln}\, \frac{1+x}{1-x} \quad.
\end{eqnarray}
The above integrals can be evaluated in a straight forward way.
In order to make a direct comparison with Eq.~(53), we
prefer to write the final expression in Minkowski space by
substituting  $p \rightarrow  \sqrt{-p^2}$  and $p^2 \rightarrow -p^2$:
\begin{eqnarray}
   F(p^2)&=& 1 - \frac{\pi \, \alpha \xi}{4 \sqrt{-p^2}}  - 
               \frac{\alpha^2 \xi^2}{4p^2}
              + \frac{3 \alpha^2}{4 p^2} \left(1+\frac{\pi^2}{12} \right)
	       - \frac{\alpha^2 \xi}{2p^2} \left(1-\frac{\pi^2}{4} 
	       \right) +{\cal O}(\alpha^3) \,.
\end{eqnarray}
One can note various important features of this result:

\begin{itemize}

\item

    $F(p^2)  \neq 1$ in the Landau gauge.

\item 

 The existence of constant and ${\cal O} (\xi)$ terms at 
 ${\cal O}(\alpha^2)$ implies the violation
 of the transversality condition. We shall elaborate more on this
  remark in Sect. 4.2.

\item 

         Eq.~(52) is derived from the LKF 
transformations based upon the assupmtion that $F=1$ in the Landau gauge.
As we have seen, this assumption is not correct to ${\cal O}(\alpha^2)$,
and therefore, Eq.~(52) is not expected to hold true in general, as
is confirmed on comparing Eq.~(53) and Eq.~(75).
However, a comparison between the two results suggests that it
contains the correct ${\cal O}(\xi^2)$ term at the level ${\cal O}(\alpha^2)$,
though it does not reproduce other terms appearing in the exact perturbative
calculation.

\end{itemize}

\subsection{Burden and Tjiang Transversality Condition}

The perturbative expression for $F(p^2)$ to the two loops
shows that the Burden-Tjiang transversality condition does not hold true
beyond one loop order. 
 Now we explicitly calculate the left hand side of Eq. (45).
 In the most general form, it can be expanded as:
 \begin{eqnarray}
   i \,  \int \frac{d^3k}{(2 \pi)^3}
  \; \gamma^{\mu} \,  S_F(k) \, \Gamma^{\nu}(k,p) \, 
  {\Delta_{\mu \nu}^0}^T(q)&=& A(p^2) + B(p^2) \not \! p \quad,
\end{eqnarray}
where the multiplication with $i$ is only for mathematical convenience.
$A(p^2)$ and $B(p^2)$ can be extracted by taking the trace of the
above equation, having multiplied by $1$ and $\not \! p$ respectively. 
With the bare fermion being
massless, it is easy to see that on doing the trace algebra and
contracting the indices, $A(p^2)=0$. Our evaluation of $F(p^2)$ helps us
identify $B(p^2)$ from Eq. (75) so that:
\begin{eqnarray}
\nonumber   && i \,  \int \frac{d^3k}{(2 \pi)^3}
  \; \gamma^{\mu} \,  S_F(k) \, \Gamma^{\nu}(k,p) \, 
  {\Delta_{\mu \nu}^0}^T(q)= \\
&& \hspace{20mm} \left[   
              - \frac{3 \alpha}{16 \pi p^2} \left(1+\frac{\pi^2}{12} \right)
	       + \frac{\alpha \xi}{8 \pi p^2} \left(1-\frac{\pi^2}{4} 
	       \right)  + {\cal O}(\alpha^2) \right] \; \not \! p  \;.
\end{eqnarray}
Obviously, for $\xi=0$,
\begin{eqnarray}
 i \,  \int \frac{d^3k}{(2 \pi)^3}
  \; \gamma^{\mu} \,  S_F(k) \, \Gamma^{\nu}(k,p) \, 
  {\Delta_{\mu \nu}^0}^T(q)  \mid_{\xi=0} &=&
            \left[   - \frac{3 \alpha}{16 \pi p^2} \left(1+\frac{\pi^2}{12} 
	        \right)+ {\cal O}(\alpha^2) \right]
	       \; \not \! p  \;,
\end{eqnarray}
which is a violation of the transversality condition at the two
loop level.

\section{Conclusions}

    In this paper, we present the one loop calculation of the
fermion-boson vertex in QED3 in an arbitrary covariant gauge for
massless fermions.  In the most general form, the vertex 
can be written in terms of 12 independent Lorentz vectors. 
Following the procedure outlined by Ball and Chiu,
4 of the 12 vectors define the longitudinal vertex.
It satisfies the Ward-Takahashi identity which relates it to the
fermion propagator. The transverse vertex is written
in terms of the remaining 8 vectors. For massless fermions, only
4 of these vectors contribute. Subtraction of the longitudinal vertex 
from the full vertex yields the transverse vertex. We evaluate the
coefficients of the basis vectors for the transverse vectors to 
${\cal O}(\alpha)$. Moreover, using this result, we calculate
$F(p^2)$ analytically to ${\cal O}(\alpha^2)$ and find that the
transversality condition does not hold true to this order. Therefore, any
non-perturbative construction of the transverse vertex based upon this
condition cannot be correct. 

    Knowing the vertex in any covariant gauge
may give us an understanding of how the essential gauge dependence of the
vertex demanded by its Landau-Khalatnikov transformation~\cite{trans,LK} is
satisfied non-perturbatively. Moreover, the perturbative knowledge of
the coefficients of the transverse vectors provides a reference for
the non-perturbative construction of the vertex as every {\em ansatz}
should reduce to this perturbative result in the weak coupling regime.
The evaluation of $F(p^2)$ to ${\cal O}(\alpha^2)$ in an arbitrary covariant
gauge
should also serve as a useful tool in the hunt for the non-perturbative
vertex which is connected to the former through Ward-Takahashi
Identity  and the Schwinger-Dyson equations. Any
vertex ansatz must reproduce Eq.~(75) for $F(p^2)$ to ${\cal O}(\alpha^2)$ 
when the coupling is weak,
leading to a more reliable 
 non-perturbative truncation of
Schwinger-Dyson equations.

\vfil\eject

\noindent{\bf{Acknowledgements}}
{}~~A.B. and A.K. are grateful for the hospitality offered to them 
by the Abdus Salam International Centre for Theoretical Physics (ICTP), 
Trieste, Italy for their stay there in the summer of 1998. M.R.P. 
wishes to thank the Special Centre for the Subatomic Structure of
Matter (CSSM), University of Adelaide, Adelaide, Australia, for the
hospitality extended by them during his visit of December 1998--February 1999 
to the Centre.

\newpage
\setcounter{section}{0}
\renewcommand{\thesection}{Appendix}
\section{}
\setcounter{equation}{0}

\baselineskip=5.5mm
\vskip 5mm
\noindent Following are the integrals used in the calculation
presented in the paper:
\begin{eqnarray}
Q_{1}&=&K^{(0)}=\int_{M}d^{3}w\, \frac{1}{(k-w)^2\,(p-w)^2}
\nonumber\\
&=& \frac{i \pi^3}{\sqrt{-q^2}} \nonumber \\ \\
Q_{2}^{\mu}&=&\int_{M}d^{3}w\,\frac{w^{\mu}}{(k-w)^2\,(p-w)^2}\nonumber\\
&=& \frac{i \pi^3}{2 \sqrt{-q^2}} \; (k+p)^{\mu} \nonumber \\ \\
Q_{3}&=&\int_{M}d^{3}w\, \frac{1}{w^4\,(k-w)^2} \nonumber\\ 
&=& 0 \nonumber \\ \\
Q_{4}&=& I^{(0)} = \int_{M}d^{3}w\, \frac{1}{w^4\,
(k-w)^2 \, (p-w)^2 }  \nonumber\\
&=& \frac{k \cdot p}{k^2 p^2} \; J^{(0)} \nonumber \\ \\
Q_{5}&=& J^{(0)}= \int_{M}d^{3}w\,\frac{1}{ w^2 \, (p-w)^2\,
(k-w)^2}\nonumber\\ \nonumber \\ 
J_0 &=& \frac{2}{i \pi^3} \, J^{(0)} \; = \; \frac{-2 \pi}{\sqrt{- k^2 p^2 q^2}}
\\  \nonumber \\ 
I_{-1} &=& \frac{2}{3kp} \; \left[ p(3k^2 + p^2) \theta(k-p) + k
            (k^2 + 3 p^2)  \theta(p-k) \right]
	    \\ \nonumber \\
I_0 &=& 2  \\ \nonumber \\
I_1 &=& \left[ \frac{2}{k} \theta(k-p) + \frac{2}{p} \theta(p-k) \right]
 \\ \nonumber \\
I_2 &=& \frac{1}{2kp} \; {\rm ln}\, \frac{(k+p)^2}{(k-p)^2}
\\ \nonumber \\
I_3 &=& \frac{2}{kp(k^2-p^2)} \left[ p \theta(k-p) - k \theta(p-k) \right]
\\ \nonumber \\
I_4 &=& \frac{2}{(k+p)^2 (k-p)^2}
%J_0 &=& \frac{1}{\Delta \sqrt{-p^2}} \; \int_0^1 \, dz \; z^{-1/2} \,
%        (z-1)^{-1/2} \, (\lambda_1 - z)^{-1} \\
%    &+& \frac{1}{\Delta \sqrt{-k^2}} \; \int_0^1 \, dz \; z^{-1/2} \,
%        (z-1)^{-1/2} \, (\lambda_2 - z)^{-1} \\
%    &-& \frac{1}{\Delta \sqrt{-q^2}} \; \int_0^1 \, dz \; z^{-1/2} \,
%        (z-1)^{-1/2} \, (\lambda_3 - z)^{-1}
\end{eqnarray}
%where
%\begin{eqnarray}
%  \lambda_1= \frac{ \Delta - k \cdot p + k^2 }{2 \Delta} \hspace{10mm}
%  \lambda_2= \frac{ \Delta - k \cdot p + p^2 }{2 \Delta} \hspace{10mm}
%  \lambda_3= \frac{-2 \Delta}{ k \cdot p - \Delta } \;.
%\end{eqnarray}
%
%
%
%
%
%
%
%
%
%
%
%

\vfil\eject

{\centerline{\large{\bf{ Figures}}}}

\vspace{0.5cm}
\epsfbox[60 50 -50 250]{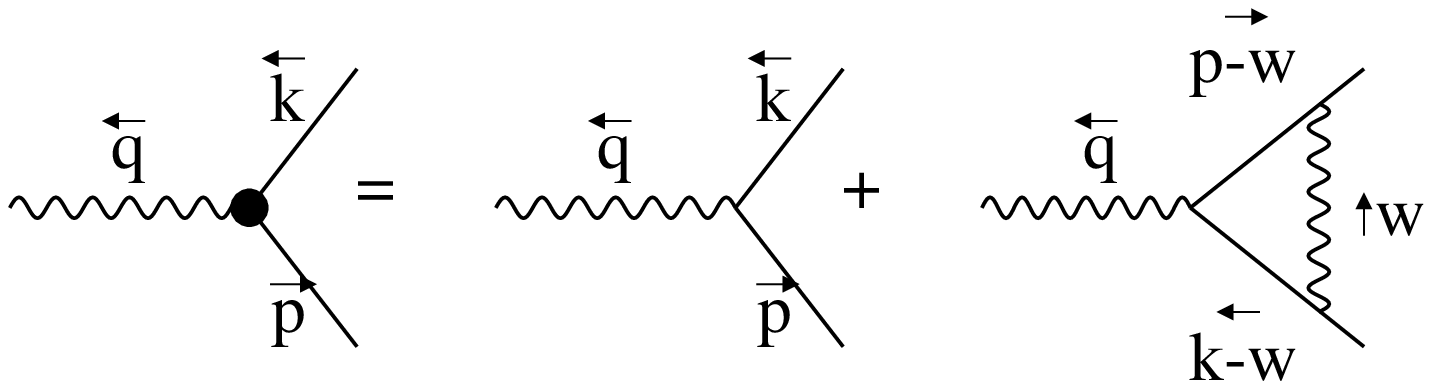}
%The first coordinate is of x-axia. A lower value of the coordinate seems
%to push the diagram towards right.

\vspace{0.5cm}
\noindent
{\hspace{45mm}{\bf Fig. 1.} One loop correction to the vertex.}

\epsfbox[90 50 -125 280]{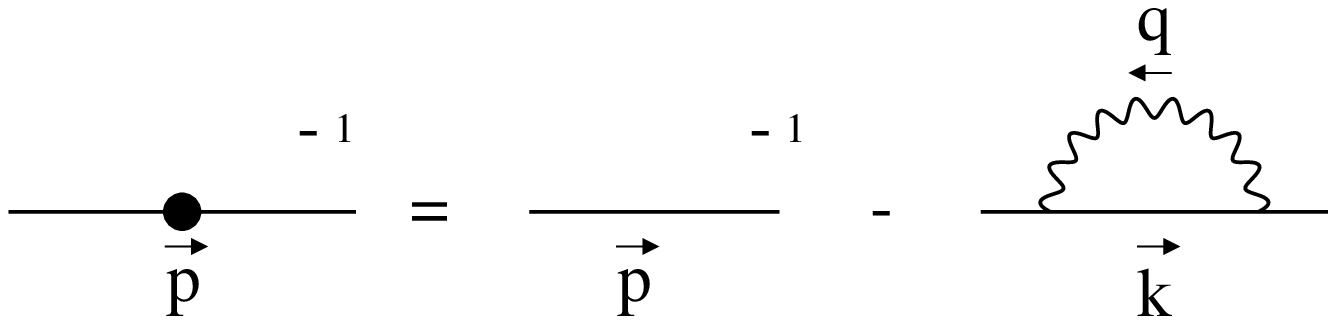}
%The 4th coordinate determines the vertical position of
%the diagram. A higher no. seems to push the diagram lower.
%The 2nd coordinate seems to contral the distance between the 
%picture and the text. The lower value seems to push the text 
%downwards.

\vskip 1cm
\noindent
{\hspace{35mm}{\bf Fig. 2.}
One loop correction to the fermion propagator.}

\end{document}